\documentclass[aps,pre,amssymb,twocolumn,superscriptaddress]{revtex4-1}
\usepackage{graphicx}
\usepackage{amsmath}
\usepackage{makecell}

\begin{document}
\title{Describing traveler choice behavior using the free utility model}
\author{Hao Wang}
\affiliation{Key Laboratory of Transport Industry of Big Data Application Technologies for Comprehensive Transport, Beijing Jiaotong University, Beijing  {\rm 100044}, China}
\affiliation{Institute of Transportation System Science and Engineering, Beijing Jiaotong University, Beijing {\rm 100044}, China}

\author{Xiao-Yong Yan}
\email{yanxy@bjtu.edu.cn}
\affiliation{Institute of Transportation System Science and Engineering, Beijing Jiaotong University, Beijing {\rm 100044}, China}

\begin{abstract}
Travel demand forecasting is an essential part of transportation planning and management. The four-step travel model is the traditional and most-common procedure utilized for travel demand forecasting, and many models have been proposed in the literature to describe each step separately. However, there is still a lack of a unified modeling framework that can successfully describe the collective choice behavior of travelers interacting with each other at different steps. This study uses the \emph{free utility model}, whose objective function is mathematically consistent with the free energy in physics, to separately and simultaneously describe travelers' mode, destination, and route choice behaviors. The free utility model's basic assumption is that the travelers will trade off the expected utility and information-processing cost to maximize their own utility. This model provides not only a unified modeling framework for traveler choice behavior, but also provides an underlying explanation for the user equilibrium model in transportation science and the potential game model in game theory.
\end{abstract}

\maketitle

\section{Introduction}\label{sec:1}
Travel is a very crucial part of humans' life \cite{Kitamura97}. When traveling, several questions arise regarding the urge to travel, the travel destination, the travel time, the right travel mode, the best travel route, and any possible travel restrictions \cite{behnew}. The collective choice behavior of travelers is the main reason for the complex traffic flow phenomenon that is observed daily. Modeling traveler choice behavior to predict travel demand continues to receive substantial attention in recent years, both in academic research and transportation planning \cite{Rasouli12,Ghader19}. The traditional travel demand forecasting involves four steps: trip generation, trip distribution, mode split, and traffic assignment (i.e., the four-step travel forecasting procedure) \cite{Wang18}. Although each step contains many alternative models, the modeling mechanisms underlying these models, within each different step, are different. For example, while predicting trip generation often uses the linear regression model, predicting trip distribution often uses the gravity model \cite{Dios11} that originally generated from an analogy with Newton's law of universal gravitation. Similarly, while predicting mode split often uses the Logit model \cite{DoMc75} that estimates the probability of choosing each alternative, predicting traffic assignment often uses the route choice game model, namely the user equilibrium (UE) model \cite{Beckmann55} that considers the interaction between travelers. In fact, the travelers' choice of trip frequency, travel time, destination, mode, and route are all human choice behaviors, and those should follow a universal law. However, the modeling framework that can uniformly describe traveler choice behavior is still lacking \cite{Wu18}.

Researchers in transportation science have long used the discrete choice model based on stochastic utility theory, represented by the Logit model \cite{DoMc75}, to forecast travel demand at different steps \cite{Zhou09}.
However, most of these models are established from the perspective of individual choice behavior, and do not take into account the interaction between travelers that are prevalent in real transportation system. In practice, travelers will consider the congestion or crowding caused by interaction with other travelers when choosing destination, travel mode, departure time and/or route.
Another type of model that can perform multi-step travel demand forecasting at the same time is the combined model \cite{Sheffi85,Yao14}, such as the combined trip distribution and traffic assignment model \cite{behnew}, the combined trip distribution, mode split and traffic assignment model \cite{Boyce83}, the combined trip generation, trip distribution, mode split and traffic assignment model \cite{Ali88}. 
Most of these models convert the multi-step combined travel demand forecasting problem into the traffic assignment problem on extended network, and then use UE class model to forecast the traffic flow [11,15].
Although the UE model takes into account the interaction between travelers, why its objective function is the sum of the integrals of the network link cost function still lacks a reasonable explanation [15,16].
Recently, Wang et al. proposed the \emph{free utility model} in which interacting travelers could choose from different destinations \cite{Wang20}. Not only does this model explain the gravity law that is widely found in various complex social systems, but it also provides a new perspective for developing a unified framework for traveler choice behavior.

In this paper, we will extend the free utility model to multiple steps of travel demand forecasting. First, we use the simplest free utility model to describe the behavior of travel mode choice. Then, we use the free utility model to describe the traveler's destination choice behavior, and derive the singly-constrained gravity model and the dual-constrained gravity model. Furthermore, we use the free utility model to describe the traveler's route choice behavior on the transportation network. Finally, we establish a combined free utility model that includes travel mode, destination, and route choice behaviors.
Using the free utility model that reflects the interaction between travelers to describe travel choice behavior not only helps us to better understand the root causes behind a variety of classic travel demand models from the perspective of human choice behavior, but also provides a more interpretable unified modeling framework for travel demand forecasting.

\section{Travel mode choice}
In the context of traveler's mode choice behavior, this study is performed in a simple transportation system with only one origin $i$ and one destination $j$, and $T$ travelers from this origin can choose the mode $k$ $(k=1,2, \dots ,K)$ to travel to the destination. The simplest system with only one traveler (i.e., $T=1$) is studied first. The utility that the travel mode $k$ can bring to the traveler is $u_{k}$, which describes the traveler satisfaction with the comfort and travel cost of the mode $k$. According utility maximization theory, the traveler will choose the travel mode with the largest utility \cite{Fishburn82}. However, in reality it is difficult for the traveler to accurately perceive the utility value of all travel modes. In this case, the traveler will choose travel mode $k$ with a probability of $p_{k}$, where $p_{k}$ depends on $u_{k}$ \cite{Luce59}. If the traveler is not aware the utilities of all travel modes at all, his or her choice behavior is that the probability of all travel modes being selected is  $p_{k}=\frac{1}{K}$, and his or her expected utility $\sum_k p_{k}u_{k}$ is the average value of the utilities of all travel modes. If the traveler wants to obtain a higher utility, he or she must master more knowledge about the utility of travel modes through information processing \cite{Marsili99}. The most natural measure of the amount of information processing is the negative information entropy ($-H$), which equals $\sum_k p_{k}\ln p_{k}$ \cite{Marsili99,Wolpert12,Ortega13}. 
Assuming that the price per unit of information is $\tau$, then the information-processing cost is $-\tau H$. At this point, the traveler must trade off the expected utility and information-processing cost \cite{Guan20, Tkacik16} to achieve the goal of maximizing total utility (max $w$). That is:
\begin{equation}
\label{eq1}
\begin{aligned}
\max w=& \sum_k p_{k}u_{k}+\tau H, \\
	\mathrm{s.t.} \quad &\sum_k p_{k}=1.
\end{aligned}
\end{equation}
Using the Lagrange multiplier method, we can obtain the following:
\begin{equation}
\label{eq2}
L(p_{k},\lambda)=\sum_k  p_{k}u_{k}+\tau H-\lambda(\sum_k p_{k}-1),
\end{equation}
where $\lambda$ is the Lagrange multiplier, and $L$ is the Lagrange function. Since $\frac{\partial L}{\partial p_{k}}=0$ for all travel modes, we can obtain the following expression:
\begin{equation}
\label{eq3}
u_{k}-\tau(\ln p_{k}+1)=\lambda.
\end{equation}
This expression means that all travel modes have the same utility \emph{minus} the marginal information-processing cost under the traveler's optimal choice strategy. This is very similar to the consumer equilibrium in microeconomics \cite{Tewari03}, where each commodity's marginal utility is equal. Therefore, the expression $u_{k}-\tau(\ln p_{k}+1)$ can be named as the marginal utility of travel mode $k$. The system's total utility is the sum of the integral of the marginal utility of all travel modes. The optimal choice strategy for travelers is to follow the \emph{equimarginal principle} \cite{Tewari03} to choose the travel mode to maximize the total utility. Combining Eq.~(\ref{eq3}) and $\sum_k p_{k}=1$ we can obtain the following expression which represents the equilibrium solution of Eq.~(\ref{eq1}):
\begin{equation}
\label{eq4}
p_{k}=\frac{\mathrm{e}^{u_{k} / \tau}}{\sum_k\mathrm{e}^{u_{k} / \tau}}.
\end{equation}
If the information-processing cost equals zero (i.e., the price per unit of information $\tau=0$), the traveler will only choose the travel mode with the largest utility. However, if the information-processing cost is very high (i.e., $\tau \to \infty$), the traveler will not consider the expected utility to choose the travel mode. In this case, the traveler will only care about the information-processing cost; thus, the traveler will choose the travel mode completely randomly. Eq.~(\ref{eq4}) has the same mathematical expression as the Logit model derived from the stochastic utility theory \cite{DoMc75}. However, the derivation presented in this study does not need to assume in advance that the utility perception error of travel mode follows the independent and identically distributed Gumbel distribution.

Furthermore, we expand the previously mentioned transportation system to a situation where there is an infinite number of travelers (i.e., $T\gg1$) with the same attributes, and each traveler has only one trip. The question then becomes how these trips are distributed among the various travel modes. In practice, the utility of the travel mode $k$ will be affected by the number of travelers using this mode (i.e., $T_{k}$); that is, $u_{k}$ is a function of $T_{k}$. For example, increasing the number of passengers on a bus will reduce the comfort of bus passengers. Therefore, the utility can be written as $u_{k}(T_{k})$, which is a monotonically decreasing and differentiable function. It is well-known that every traveler follows the equimarginal principle to make the best choice. In this case, one traveler's choice of travel mode depends on how other travelers are distributed under all travel modes \cite{Conlisk76}. This phenomenon in which an individual's behavior depends on other individuals' behaviors, called individual interaction, is very common in social systems. In this interactive system, the travelers' optimal choice strategy is to make all travel modes have the same marginal utility. Since the total utility of the system is the sum of the integral of the marginal utility of all travel modes, the utility maximization model of the interactive system can be written as follows:
\begin{equation}
\label{eq7}
\begin{aligned}
\max W=& \sum_k \int_0^{T_{k}} u_{k}(x) \mathrm{d} x+\tau S, \\
	\mathrm{s.t.} \quad &\sum_k T_{k}=T,
    \end{aligned}
\end{equation}
where $S$ is the sum of the information entropy $H$ of each trip (i.e., $S=T \cdot H$). When $\tau=0$, Eq.~(\ref{eq7}) becomes equivalent to a potential game with an infinite number of players \cite{Monderer96}. When $u_k$ is constant (i.e., there is no interaction between travelers), Eq.~(\ref{eq7}) degenerates into a Logit model whose solution is $T_{k}=T\frac{e^{u_k / \tau}}{\sum_k e^{u_k / \tau}}$.

The objective function of Eq.~(\ref{eq7}) is mathematically consistent with the Helmholtz free energy ($F$) in physics \cite{Kittel80}. Therefore, Eq.~(\ref{eq7}) can be named as \emph{free utility model} \cite{Wang20}. In other words, this travel mode choice system can be analogized to an isothermal and isochoric thermodynamic system that contains several subsystems and is in thermal contact with the thermal reservoir. That is, the number of travelers is analogous to the number of particles; the first term of the objective function of Eq.~(\ref{eq7}) is analogous to the thermodynamic system's internal energy containing potential energy (therefore Monderer and Shapley named this item potential function \cite{Monderer96}); the information-processing price is analogous to the temperature of the thermal reservoir; the information entropy is analogous to the entropy of the thermodynamic system; the information-processing cost is analogous to the heat exchanged between the thermodynamic system and the thermal reservoir; and the marginal utility is analogous to the chemical potential of the subsystem. However, the essence of these two systems is different: that is, while the thermodynamic system follows the minimum free energy principle to bring the system to equilibrium (in which all subsystems have the same chemical potential), the maximization of the free utility in the travel mode choice system is the result of travelers following the equimarginal principle to make choices to maximize their utility.

\section{Travel destination choice}
We further use the free utility model to describe the destination choice behavior of travelers in a transportation system including $M$ origins labeled $i$ $(i=1, 2, \dots , M)$ and $N$ destinations labeled $j$ $(j=1, 2, \dots , N)$. $O_{i}$ is the number of travelers from origin $i$, $T_{ij}$ is the number of travelers from the origin $i$ to the destination $j$, and $u_{ij}(T_{ij})$ is a monotonically decreasing and differentiable function of $T_{ij}$ that describes the attractiveness of the destination $j$ to travelers and the satisfaction of travelers with the travel cost from the origin $i$ to the destination $j$. Similar to Eq. ~(\ref{eq7}), the free utility model of this system can be written as follow:
\begin{equation}
\label{eq77}
\begin{aligned}
\max W=& \sum_j \int_0^{T_{ij}} u_{ij}(x) \mathrm{d} x+\tau S_i, \\
\mathrm{s.t.} \quad &\sum_j T_{ij}=O_{i},
\end{aligned}
\end{equation}
where $S_i=-\sum_j T_{ij}\ln \frac{T_{ij}}{O_i}$ represents the information-processing entropy of $O_i$ travelers.

The free utility model in Eq.~(\ref{eq77}) can be directly used  to forecast the trip distribution after setting the appropriate utility function $u_{ij}(T_{ij})$ according to the real data.
In addition, the free utility model can also be used to derive the gravity model that is widely used in trip distribution forecasting.
In order to derive the classic gravity model, the utility function needs to be the logarithmic relationship of the number of travelers from the origin $i$ to the destination $j$, i.e., $u_{ij}(T_{ij})=A_{j}-c_{ij}-\gamma \ln T_{ij}$, where $A_j$ is the constant attractiveness of the destination $j$, which reflects the activity opportunities (e.g., retail activities and employment density) of the destination \cite{Sheffi85}, $c_{ij}$ is the constant travel cost from $i$ to $j$, $\gamma \ln T_{ij}$ is the travel congestion cost function, and $\gamma$ is a non-negative parameter. Equation~(\ref{eq77}) can be specifically written as follows:
\begin{equation}
\label{eq8}
\begin{aligned}
\max W=& \sum_j \int_0^{T_{ij}} (A_{j}-c_{ij}-\gamma \ln x) \mathrm{d} x+\tau S_i, \\
	\mathrm{s.t.} \quad &\sum_j T_{ij}=O_{i}.
    \end{aligned}
\end{equation}
Using the Lagrange multiplier method, we can obtain the following:
\begin{equation}
\label{eq9}
T_{ij}=O_{i}\frac{\mathrm{e}^{(A_j-c_{ij}) / (\gamma+\tau)}}{\sum_j \mathrm{e}^{(A_j-c_{ij}) / (\gamma+\tau)}}.
\end{equation}
This expression is the singly-constrained gravity model \cite{Dios11}. The parameter $\tau$ reflects the travelers' information-processing price, and $\gamma$ reflects the intensity of interaction between the travelers. 
When $\gamma > 0$ and $\tau = 0$ (i.e., no information-processing cost), the free utility model is equivalent to the degenerated DCG model established with the potential game theory \cite{Yan19}.
When $\gamma >0$ and $\tau \to \infty$ (i.e., very high information-processing cost), travelers can only uniformly and randomly choose destination.
When $\tau > 0$ and $\gamma = 0$ (i.e., no interaction between travelers), the free utility model becomes equivalent to the Logit model \cite{DoMc75}.
Finally, when $\tau = 0$ and $\gamma = 0$, all travelers will choose the destination with the highest utility.

The most used trip distribution model in transportation science is the dual-constrained gravity model \cite{Dios11}, which satisfies two conditions: $\sum_j T_{ij}=O_i$ and $\sum_i T_{ij}=D_j$, where $D_j$ is the number of travelers attracted by the destination $j$. If $D_j$ is constant in this system, the free utility model can be extended as follows:
\begin{equation}
\label{eq10}
\begin{aligned}
\max W=&  \sum_i \sum_j \int_0^{T_{ij}} (A_{j}-c_{ij}-\gamma \ln x) \mathrm{d} x+\tau  \sum_i S_i, \\
	\mathrm{s.t.} \quad &\sum_j T_{ij}=O_{i},\\
	&\sum_i T_{ij}=D_{j}.
    \end{aligned}
\end{equation}
The free utility model solution is the doubly-constrained gravity model, as follows:
\begin{equation}
\label{eq11}
T_{ij}=a_{i}b_{j}O_{i}D_{j}\mathrm{e}^{-c_{ij} / (\gamma+\tau)},
\end{equation}
where $a_{i}=1/\sum_j b_j D_j \mathrm{e}^{-c_{ij} / (\gamma+\tau)}$ and $b_{j}=1/\sum_i a_i O_i \mathrm{e}^{-c_{ij} / (\gamma+\tau)}$ are balancing factors.

Under specific parameter combinations, Eq.~(\ref{eq10}) can be reduced to some classic models. 
When $\tau > 0$ and $\gamma=0$ (i.e., no interaction between travelers), Eq.~(\ref{eq10}) becomes equivalent to the free cost model proposed by Tomlin et al. \cite{Tomlin68}, and the solution of Eq.~(\ref{eq10}) is the same as that of Wilson's maximum entropy model \cite{Wilson67}.
When $\tau = 0$ and $\gamma = 0$, Eq.~(\ref{eq10}) becomes equivalent to the Hitchcock-Koopmans problem \cite{Hitchcock41}, which asks how many travelers from $i$ to $j$ for each pair of origin and destination can minimize the total cost $\sum_i \sum_j (c_{ij}-A_{j})T_{ij}$.
When $\tau \to \infty$ (i.e., the first term of Eq.~(\ref{eq10}) can be ignored), Eq.~(\ref{eq10}) becomes equivalent to Sasaki's prior probability model whose solution is $T_{ij}\propto O_iD_j$ \cite{Tomlin68}.

\section{Travel route choice}
Here, we use the free utility model to describe the travelers' route choice behavior in a transportation network, in which there are $M$ origins labeled $i$ $(i=1, 2, \dots , M)$, $N$ destinations labeled $j$ $(j=1, 2, \dots , N)$, and $L$ links labeled $a$ $(a=1, 2, \dots , L)$. The finite number of travelers from the origin $i$ to the destination $j$ is $T_{ij}$. Travelers have one or more alternative routes from $i$ to $j$. Each link has a utility function $u_a(x_a) = -t_a(x_a)$, which reflects the relationship between the link cost $t_a$ and the link flow $x_a$. The free utility model of this network can be written as follows:
\begin{equation}
\label{eq13}
\begin{aligned}
\max W=&  -\sum_a \int_0^{x_a} t_a(x) \mathrm{d} x +\tau S, \\
\mathrm{s.t.} \quad &\sum_k f^k_{ij}=T_{ij},\\
&f^k_{ij}\geq0,
\end{aligned}
\end{equation}
where $S=-\sum_{i}\sum_{j}\sum_{k} f^k_{ij} \ln \frac{f^k_{ij}}{T_{ij}}$ is the information-processing entropy for all travelers, and $f^k_{ij}$ is the flow on the $k^{th}$ route from $i$ to $j$.
The relationship between the route flow $f^k_{ij}$ and the link flow $x_a$ in the transportation network is $x_a=\sum_i\sum_j \sum_k f^k_{ij} \delta^{a,k}_{ij}$, where $\delta^{a,k}_{ij}$ is a switch parameter with a binary value. If the route $k$ includes the link $a$, $\delta^{a,k}_{ij}$ equals 1; otherwise, it equals 0.

From Eq.~(\ref{eq13}), we can see that the transportation network's free utility model is mathematically consistent with the stochastic user equilibrium (SUE) model \cite{Fisk80} used for traffic assignment in transportation science. There are many algorithms that can solve this model and calculate the flow on all links within the network \cite{Sheffi85}. In the case where the information-processing price equals zero (i.e., $\tau=0$), the free utility model in Eq.~(\ref{eq13}) is mathematically consistent with the classic user equilibrium (UE) model established by Beckmann \cite{Beckmann55}. The equilibrium solution of the UE model, where all routes used between each origin-destination (OD) pair have equal and minimum costs, is exactly the result of travelers' optimal route choice strategy following the equimarginal principle. The UE model's objective function, i.e., the sum of the integral of the marginal utility (negative cost) of each link, is a negative free utility without information-processing cost. This provides a new perspective for understanding the UE and SUE models in transportation science.

\section{Combined model}
Up to now, we have used the free utility model to separately describe the travelers' mode choice behavior, destination choice behavior and route choice behavior. In practice, however, travelers do not make the above choices in separate stages, but consider the travel destination, mode of transport, and travel route simultaneously. Therefore, we need a combined model for these traveler choice behaviors under the free utility framework.

\begin{figure}
	\centering
	\includegraphics[width=0.75\columnwidth]{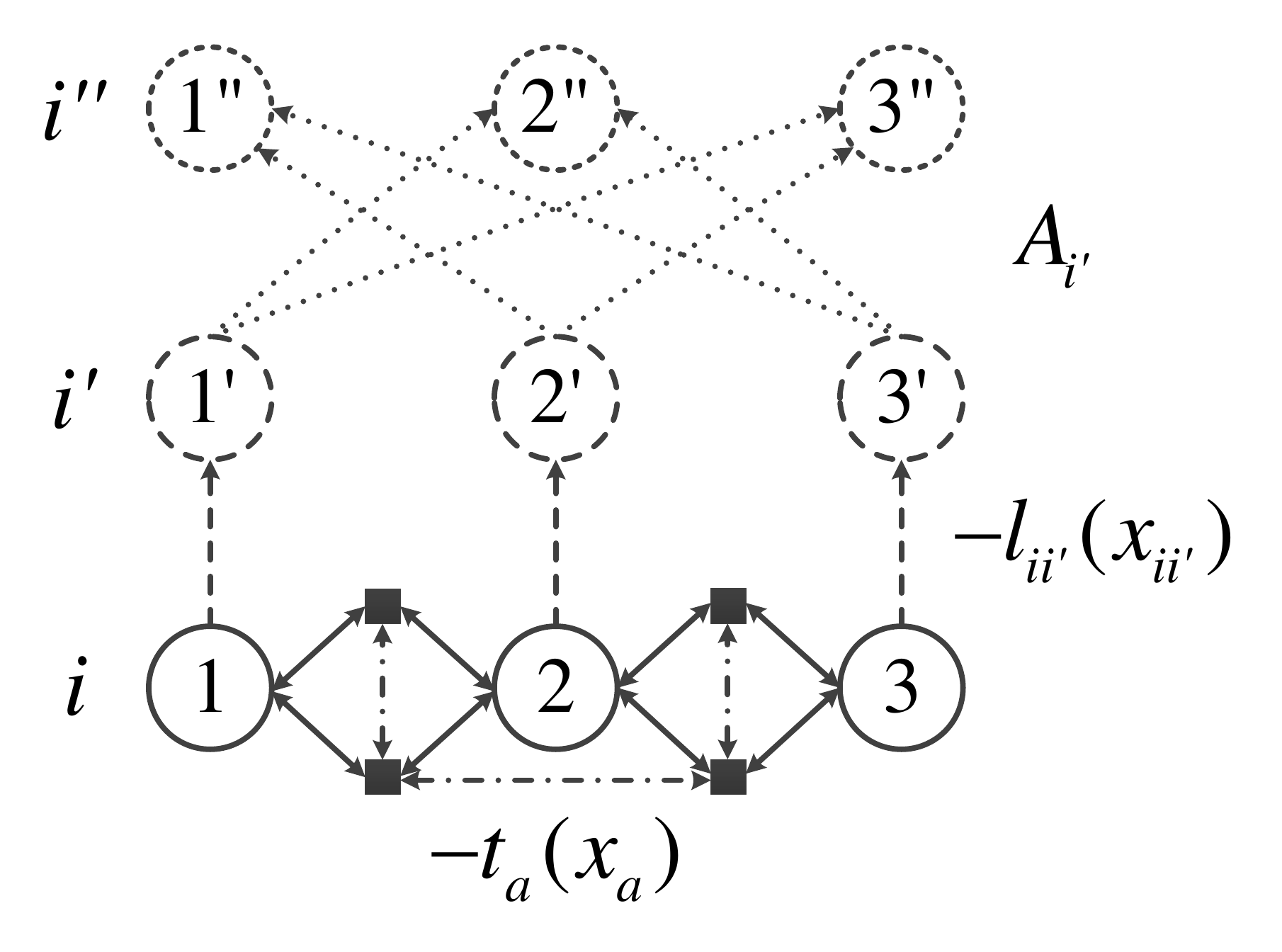}
	\caption{{\bf An expanded transportation network.}  }
	\label{fig1}
\end{figure}

In this study, we present a network expansion method to simultaneously solve the travel mode, destination and route choice behaviors, as shown in Fig.~\ref{fig1}.
The bottom layer is the real multi-mode transportation network, in which each circle node labeled $i$ represents a location (which is the centroid of the traffic analysis zone (TAZ) in transportation science) that can produce and attract trips, the square nodes represent ordinary nodes such as intersections or subway stations, and the links labeled $a$ represent segments. The solid lines between the nodes are the road links, and the dash-dot lines are the subway links. Each link is assigned a utility function $u_a(x_a)$ summarizing the relationship between link utility $u_a$ and link flow $x_a$.
We expand the real transportation network by adding two types of dummy nodes and two types of dummy links. First, we add a dummy destination node $i'$ for each node $i$ and a dummy link $ii'$ from node $i$ to node $i'$ (see the circle nodes in the middle layer and the dashed lines from the bottom layer to the middle layer in Fig.~\ref{fig1}). 
The flow along link $ii'$ is $x_{ii'}$, which is equal to the trip attraction at the real destination $i$. The equivalent utility of link $ii'$ is the negative destination variable attractiveness $-l_{ii'}(x_{ii'})$. Next, we add a dummy origin node $i''$ for each node $i$ and add dummy links from node $i'$ to all dummy origin nodes except the corresponding dummy origin $i''$ (see the circle nodes in the top layer and the dotted lines from the middle layer to the top layer in Fig.~\ref{fig1}). The flow along link $i'i''$ is $x_{i'i''}$, which is equal to the number of trips between the corresponding origin and destination pair. The equivalent utility of link $i'i''$ is the constant attractiveness $A_{i'}$. In this manner, an expanded network is formed.

In this expanded network, the fixed trip volume from origin node $i$ to corresponding dummy origin node $i''$ is $O_i$, i.e., the number of travelers at origin $i$. Consequently, the combined travel mode, destination and route choice problem is converted into a route choice problem in the expanded network. The free utility model for this expanded network can be written as follow:
\begin{equation}
\label{eq14}
\begin{aligned}
\max W=&  \sum_a \int_0^{x_a} u_a(x) \mathrm{d} x-\sum_{i'} \int_{0}^{x_{ii'}} l_{ii'}(x) \mathrm{d} x\\
&+\sum_{i''}\sum_{i'\ne i''} A_{i'} x_{i'i''}+\tau  S, \\
\mathrm{s.t.} \quad &\sum_k f^k_i=O_{i},\\
&f^k_i\geq0,
\end{aligned}
\end{equation}
where $S= -\sum_i \sum_{k} f^k_i \ln \frac{f^k_i}{O_i}$ is the information-processing entropy for all travelers, and $f^k_i$ is the flow on the $k^{th}$th route from node $i$ to the corresponding dummy origin node $i''$.
The relationship between the route flow $f^k_i$ and the link flow $x_a$ in the expanded network is $x_a=\sum_i\sum_k f^k_i \delta^{a,k}_i$, where $\delta^{a,k}_i$ equals 1 if route $k$ uses the link $a$; otherwise, it equals 0. 
Similar to Eq.~(\ref{eq13}), Eq.~(\ref{eq14}) is a SUE model for traffic assignment on the expanded network.
By solving it, we can obtain not only the traffic or passenger flow on the link of the real multi-mode transportation network but also the trip volume between TAZs (i.e., flow $x_{i'i''}$ along dummy link $i'i''$) and the trip volume attracted by each TAZ (i.e., flow $x_{ii'}$ along dummy link $ii'$). 
Such a model,  which can simultaneously predict trip distribution and traffic or passenger flow, is called a combined distribution-assignment model in transportation science \cite{Sheffi85,Yao14}. However, the combined model established by the free utility framework is more interpretable than the other combined distribution-assignment models presented in the literature. 
It clearly shows the interaction of travelers on the links and destinations of the real multi-mode transportation network.

\begin{figure}
	\centering
	\includegraphics[width=0.7\columnwidth]{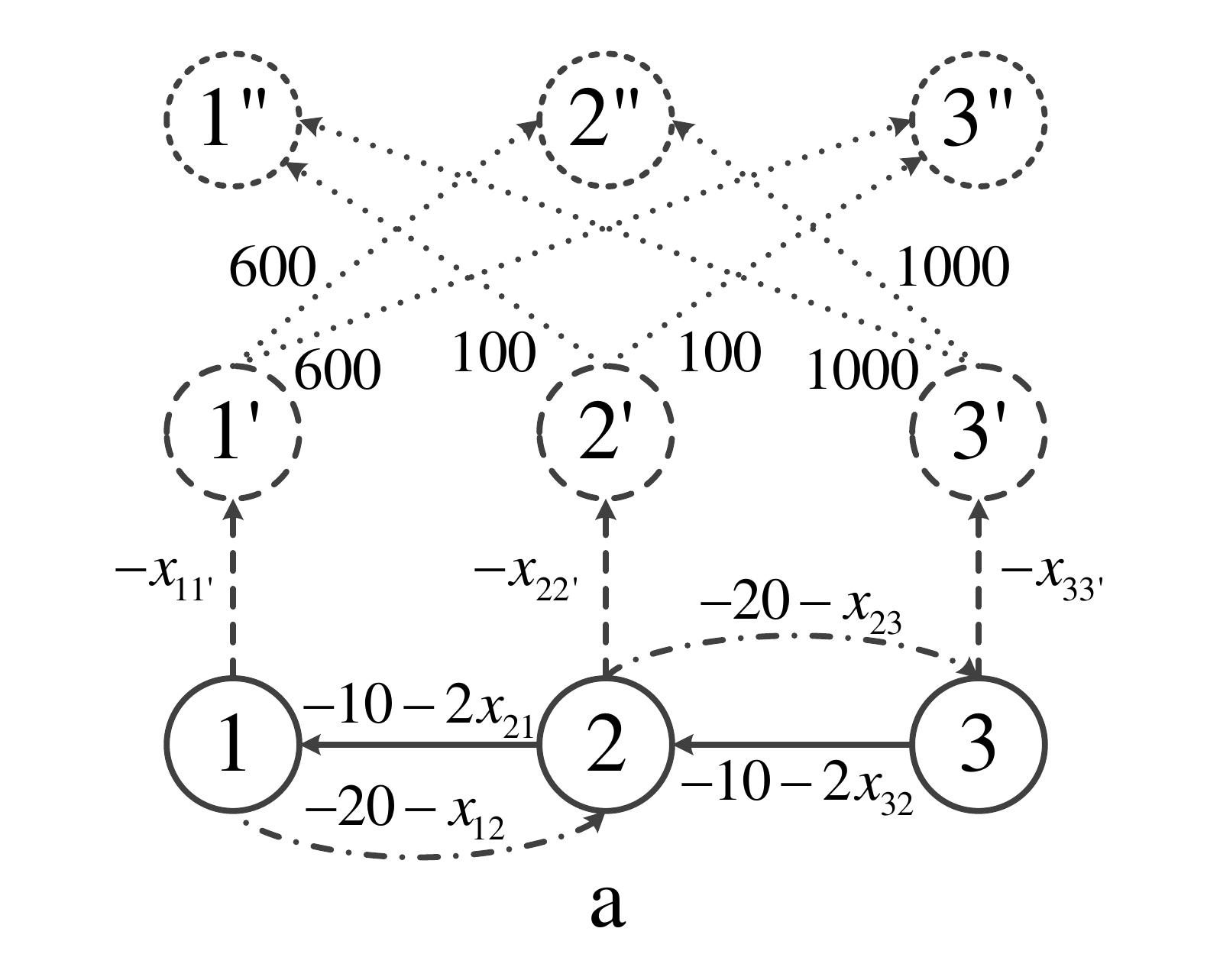}
	\includegraphics[width=0.7\columnwidth]{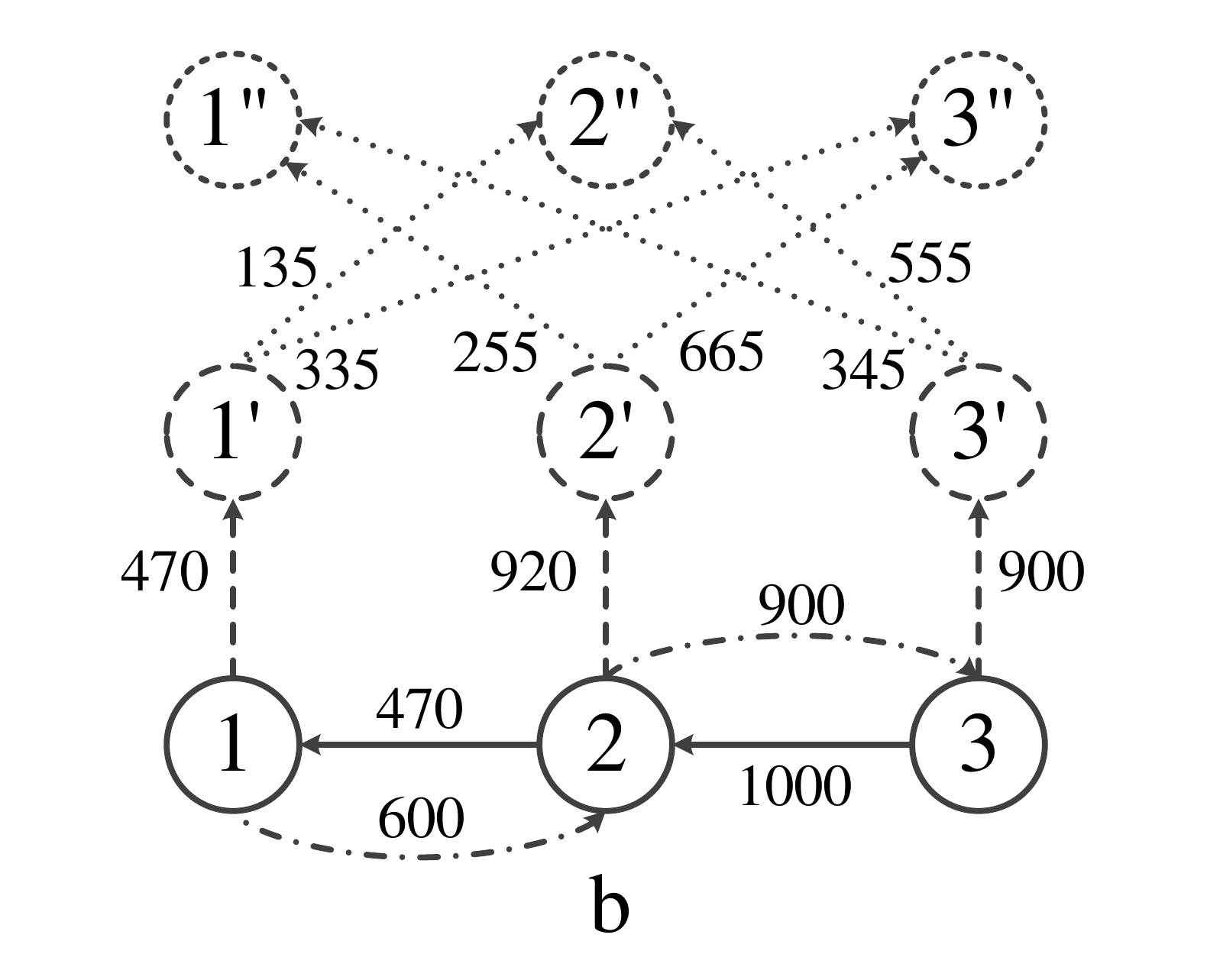}
	\caption{{\bf An example of combined model.} 
		 a. Link utility functions of a simple expanded network. b. The optimal solution of the combined model. }
	\label{fig2}
\end{figure}

To better illustrate the above combined model, we add an example here. We built a simple expanded transportation network (see Fig.~\ref{fig2}a), in which the circle nodes represent three TAZs, and their trip productions are $O_1=600$, $O_2=690$, and $O_3=1000$ respectively.
The two solid lines in the bottom layer represent road links, and the two dash-dot lines represent subway links. Their utility functions are $u_{12}=-20-x_{12}$, $u_{23}=-20-x_{23}$, $u_{32}=-10-2x_{32}$, and $u_{21}=-10-2x_{21}$ respectively, where $x$ is the link flow, the same below. 
The utility functions of the dummy links between the bottom layer and the middle layer are $u_{11'}=-x_{11'}$, $u_{22'}=-x_{22'}$, and $u_{33'}=-x_{33'}$ respectively.
The utility functions of the dummy links between the middle layer and the top layer are $u_{1'2''}=u_{1'3''}=600$, $u_{2'1''}=u_{2'3''}=100$, and $u_{3'1''}=u_{3'2''}=1000$ respectively.
To simplify the calculation, we assume that the information-processing price $\tau = 0 $. 
If the trip production of the TAZs is regarded as the trip distribution from the bottom layer nodes to the top layer nodes,
i.e., $T_{11''}=O_1=600$, $T_{22''}=O_2=690$ and $T_{33''}=O_3=1000$,
 the optimal solution of this example can be obtained by the algorithm for solving the UE model, as shown in Fig.~\ref{fig2}b. 
The link flows in the optimal solution are $x_{21}=470$, $x_{12}=600$, $x_{23}=900$, and $x_{32}=1000$ respectively. The trip attractions of TAZs are $D_1=x_{11'}=470$, $D_2=x_{22'}=920$, and $D_3=x_{33'}=900$ respectively. The trips between the TAZs are $T_{12}=x_{2'1''}=255$, $T_{13}=x_{3'1''}=345$, $T_{21}=x_{1'2''}=135$, $T_{23}=x_{3'2''}=555$, $T_{31}=x_{1'3''}=335$, and $T_{32}=x_{2'3''}=665$ respectively.
Table \ref{tab1} shows the marginal utilities of the six routes from the bottom layer nodes to the top layer nodes.

\begin{table}
	\caption{The marginal utility of the route between TAZs}
	\renewcommand{\arraystretch}{1.0}
	\begin{tabular}{p{1cm}<{\centering}p{3cm}<{\centering}lr}
		\hline
		origin & 
		destination & route & marginal utility \\ \hline
		1              
		& 2         
		& $1$-$2$-$2'$-$1''$
		& -1440
		\\
		 1              
		& 3        
		& $1$-$2$-$3$-$3'$-$1''$
		& 
		-1440                                  
		\\
		 2               
		& 1        
		& $2$-$1$-$1'$-$2''$      
		& 
		-820                                  
		\\
		2              
		& 3         
		& $2$-$3$-$3'$-$2''$             
		& 		-820 
		\\  
		 3               
		 & 1        
		 & $3$-$2$-$1$-$1'$-$3''$      
		 & 
		 -2830                                 
		 \\
		 3              
		 & 2         
		 & $3$-$2$-$2'$-$3''$
		 & -2830
		 \\  
		\hline    
	\end{tabular}
	\label{tab1}
\end{table}

\section{Discussion and conclusions}\label{sec:3}
In this paper, we separately described the travel mode, destination, and route choice behaviors of travelers using the free utility model. The free utility model has two basic assumptions: the traveler pursues maximum utility and needs to pay information-processing cost to master more knowledge about the utility of alternative \cite{Wang20}. Compared with the most used models in the four-step travel forecasting procedure, the free utility model can better describe traveler choice behavior. For example, the Logit model for predicting the probability of travel mode choice \cite{DoMc75}, the maximum entropy model \cite{Wilson67}, and the free cost model \cite{Tomlin68} for predicting trip distribution all do not reflect the interaction between travelers. While the traditional SUE \cite{Fisk80} and UE \cite{Beckmann55} models are mathematically consistent with the free utility model in describing travel route choice, the integral term of the link cost function in their objective functions still lacks a reasonable explanation. However, the free utility model can simultaneously reflect the trade-off between the expected utility and the information-processing cost as well as the interaction between travelers. Not only does this model allow us to better understand the root causes behind the aforementioned classic travel demand forecasting models from the perspective of human choice behavior, but also allows us to study the travel decision-making process of travelers in a more interpretable theoretical framework.

We further used the free utility model to establish a combined model that simultaneously considers travel mode, destination, and route choice behaviors. The combined model is more aligned with the real traveler choice behavior when compared with the traditional four-step travel forecasting procedure. Although this model only combined the travel mode, destination, and route choice behaviors in this paper, other traveler choice behaviors can also be analyzed under the same free utility framework. For example, an elastic demand link from the origin node $i$ and its corresponding dummy origin node $i''$ in Fig.~\ref{fig1} can be added, and an appropriate utility function for this link can be placed. Then, the actual number of trips generated by the origin node $i$ under the premise of a given potential maximum number of trips can be predicted. Furthermore, the free utility model can be used to model the combination of departure time choice behavior \cite{Vickrey69} and other traveler choice behaviors in a multi-state super network \cite{Fu14,Liu15}. In short, such combined models allow the systematic study of the entire travel decision-making process of travelers under the more interpretable free utility framework. Furthermore, the increasingly rich big data capturing human mobility allow for more explorations into the rules of traveler choice behavior to reasonably assign various utility functions in the free utility model, which can provide a more potential alternative tool for the quantitative analysis of complex transportation system that is full of interaction between travelers.

Although the free utility model was used in this paper to describe the traveler's choice behavior in the transportation system, this model can also be applied to many other social systems. From Eq.~(\ref{eq7}), it is clear that the free utility model is mathematically consistent with the stochastic potential game model \cite{Goeree99}, and can be reduced to the classic potential game model \cite{Monderer96}. These potential game models have been widely used to capture the players' alternative choice behaviors in many social systems. The objective function in the potential game model proposed by Monderer and Shapley \cite{Monderer96} is analogous to the potential function in physics. The variation in a player's individual payoff due to changes in the player's strategy is equal to the variation in the potential function. The equilibrium solution of the potential game is the solution when the potential function is maximized. At this time, all players have the same payoffs, and they cannot unilaterally change their strategy to increase their own payoffs. However, Monderer and Shapley raised the question ``What do the firms try to jointly maximize?" on how to explain the potential function. In fact, the potential function is not a common goal for players. From the free utility model, it has become evident that players will only maximize their own utility (i.e., the payoff), and the best strategy is to make the marginal utilities of all their alternatives equal. When all players follow this equimarginal principle, the sum of the integral of each alternative's marginal utility (called the total utility in economics) is naturally maximized. In other words, the maximum potential function is the result of each player's optimal choice strategy, but not the players' common objective. This provides an essential understanding of the potential game model and allows better modeling for the choice behaviors in various complex social systems. With all things considered, the free utility model not only helps to deeply understand the traveler choice behavior in a transportation system, but also has potential application value in predicting, guiding, and even controlling human choice behavior in various complex social systems.

{\bf Acknowledgments:} This work was supported by the National Natural Science Foundation of China (Grant Nos. 71822102, 71621001, 71871010).

\end{document}